%% file: main.tex
\UseRawInputEncoding

\documentclass[conference]{IEEEtran}

\input{sections/Preamble}
\graphicspath{{./figs/}}
\begin{document}
\bibliographystyle{unsrt}
\input{sections/title}
\input{sections/introduction}

\input{sections/preliminary}

\input{sections/motivation}

\input{sections/simulations}
\input{sections/Conclusions}

%\input{sections/Ackowledgment}

\input{references/main.bbl}
%\bibliographystyle{references/IEEEtran}
%\bibliography{references/third_ref.bib}
%\bibliography{references/new_ref.bib}
\end{document}

%% file: sections/Preamble.tex
\IEEEoverridecommandlockouts
\usepackage{cite}
\usepackage{amsmath,amsfonts,amsthm,bm,amssymb} %
\usepackage[table,xcdraw]{xcolor}
\usepackage{algorithm}
\usepackage{algpseudocode} 
\usepackage{graphicx}
\usepackage{graphics}
\usepackage{epsfig}
\usepackage{footnote}
\usepackage{textcomp}
\PassOptionsToPackage{no-math}{fontspec}
\usepackage{xcolor}
\usepackage{tikz}  
\usepackage{makecell}
\usepackage{bm}
\usetikzlibrary{calc} 
\usetikzlibrary{arrows,shapes,chains} 
\usetikzlibrary{positioning, shapes.geometric}
\usepackage{multirow} %multirow for format of table
\usepackage{verbatim}
\usepackage{enumitem}
\def\BibTeX{{\rm B\kern-.05em{\sc i\kern-.025em b}\kern-.08em
		T\kern-.1667em\lower.7ex\hbox{E}\kern-.125emX}}
	
\usepackage{verbatim}
\usepackage{pgfplots}
\usepackage{hyperref}
\pgfplotsset{width=10cm,compat=1.9}
\pgfplotsset{
	every axis legend/.style={
		cells={anchor=center},
		inner xsep=3pt,inner ysep=2pt,
		nodes={inner sep=2pt,text depth=0.1em},
		anchor=north east,
		shape=rectangle,
		fill=white,draw=green,
		font=\footnotesize%\tiny
	},
}
\usepackage{verbatim}
\usepackage{booktabs}
\usepackage{fancyhdr}
\usepackage{svg}
\normalsize
	

%% file: sections/title.tex
\title{The impact when neural min-sum variant meets ordered statistics decoding of LDPC codes\\
%	\footnotesize 
%	\thanks{applicable funding agency here. If none, delete this.}
}
\begin{comment}
\author{\IEEEauthorblockN{Guangwen Li}
	\IEEEauthorblockA{\textit{College of Information \& Electronics} \\
		\textit{Shandong Technology and Business University}\\
		Yantai, China \\
		lgwa@sdu.edu.cn}
	\and
	\IEEEauthorblockN{Xiao Yu}
	\IEEEauthorblockA{\textit{Department of Physical Sports} \\
		\textit{Binzhou Medical University}\\
		Yantai, China \\
		yuxiao@bzmu.edu.cn}
}
\end{comment}

\author{Guangwen Li, Xiao Yu

\thanks{G.Li is with the College of Information \& Electronics, Shandong Technology and Business University, Yantai, China e-mail: lgw.frank@gmail.com}% <-this % stops a space
\thanks{X.Yu is with the Department of Physical Sports, Binzhou Medical University, Yantai, China e-mail: yuxiao2020phu@gmail.com}% <-this % 
}

%\markboth{IEEE Transactions on Communications}%
%{Submitted paper}
\maketitle
\begin{abstract}
This paper introduces three key initiatives in the pursuit of a hybrid decoding framework characterized by superior decoding performance, high throughput, low complexity, and independence from channel noise variance. 

Firstly, adopting a graphical neural network perspective, we propose a design methodology for a family of neural min-sum variants. Our exploration delves into the frame error rates associated with different decoding variants and the consequential impact of decoding failures on subsequent ordered statistics decoding. Notably, these neural min-sum variants exhibit generally indistinguishable performance, hence the simplest member is chosen as the constituent of the hybrid decoding.

Secondly, to address computational complexities arising from exhaustive searches for authentic error patterns in cases of decoding failure, two alternatives for ordered statistics decoding implementation are proposed. The first approach involves uniformly grouping test error patterns, while the second scheme dynamically generates qualified searching test error patterns with varied sizes for each group. In both methods, group priorities are determined empirically.

Thirdly, iteration diversity is highlighted in the case of LDPC codes requiring high maximum iterations of decoding. This is achieved by segmenting the long iterative decoding trajectory of a decoding failure into shorter segments, which are then independently fed to small models to enhance the chances of acquiring the authentic error pattern.

These ideas are substantiated through extensive simulation results covering the codes with block lengths ranging from one hundred to several hundreds.
\end{abstract}


%\IEEEpeerreviewmaketitle

%% file: sections/introduction.tex
\section{Introduction}
Since Shannon's groundbreaking work in information theory \cite{shannon1948mathematical}, channel coding has remained fundamental in modern reliable telecommunication. Low-density parity-check codes (LDPC), within the diverse family of linear block codes, distinguish themselves by demonstrating exceptional error correction capabilities and the potential to asymptotically approach the Shannon limit \cite{gallager62, mackay96}. In practical scenarios, such as fifth-generation (5G) New Radio (NR) shared channels \cite{nguyen2019efficient} and the communication networks of the next-generation Internet of Things (IoT) \cite{yang2022design}, the pursuit of a notable frame error rate (FER), high throughput, low latency, and low decoding complexity continues to capture the attention of the coding community.

Belief propagation (BP) methods, once considered optimal for theoretically infinite-length codes, initially dominated LDPC decoding \cite{chen2005reduced}. However, their potential computational complexity poses challenges under certain stringent conditions, and their sub-optimality arises in the presence of cycles in the structures of finite-length practical codes. Despite the effectiveness of approximations like min-sum (MS) variants \cite{zhao05,jiang06} in reducing BP complexity, they commonly suffer from non-negligible performance loss. Extensive efforts to strike a better trade-off among competing objectives, such as performance and complexity, have been extensively reported in the literature \cite{li2021high,ferraz2021survey,nadal2021parallel,liu2020efficient}.

In contrast, ordered statistics decoding (OSD) \cite{Fossorier1995,Fossorier1999,Fossorier2001}, initially proposed to approximate maximum likelihood (ML) decoding for short linear block codes, was later extended to narrow the performance gap between BP and ML decoding for LDPC codes. Various strategies to accelerate OSD decoding for classical BCH codes or short LDPC codes have been introduced recently \cite{Yue2021,Yue2022}, aiming to increase decoding throughput by triggering dedicated early stopping criteria or reducing the number of test error patterns (TEPs) via thresholding to allow for the effective selection of the most likely estimate. However, these approaches require the precondition of knowing the channel noise variance, and any deviated estimation of it leads to performance degradation. Moreover, the bottleneck of the inherent serial decoding nature is hardly relieved when the OSD plays the solo role of decoding without aid from other methods.

The past decade has witnessed the integration of deep learning techniques into various domains, from image recognition \cite{razzak18} to natural language processing \cite{young18} and autonomous driving \cite{grigorescu20}. Notably, for tasks involving graph-structured data, graph neural networks (GNNs) \cite{sanchez-lengeling2021a} have demonstrated advantages over other architectures like fully connected networks (FCNs) or convolutional neural networks (CNNs) in terms of model size \cite{chollet2015keras}.

The advent of deep learning has brought new perspectives to error correction coding as well. Nachmani et al. \cite{nachmani16} pioneered the unrolling of iterative BP decoding into a neural network (NN), giving rise to neural belief propagation (NBP). This innovative approach involved associating a trainable parameter with each edge to incorporate domain knowledge of coding. The NBP concept subsequently evolved into various neural min-sum (NMS) variants. Further exploration into diverse NN architectures, including CNNs and recurrent neural networks (RNNs) \cite{nachmani18, liang18, lugosch18, wang20}, aimed to further enhance decoding performance.
In a recent study \cite{buchberger20}, the idea of generating a pruned NN by removing unimportant check nodes from the overcomplete parity check matrix was proposed. However, obtaining such a matrix remains a challenging task. Another approach presented in \cite{buchberger21} introduced a two-stage decimation process: the likely codeword bits were first decimated using an NN, followed by a list decoding of NBP to improve decoding. Nevertheless, this method still lags significantly behind the ML curve. Rosseel et al. \cite{Rosseel2022} proposed a collaborative solution involving NBP and tailored RNN models to address decoding failures caused by the presence of absorbing sets or trapping sets inherited in the code. Their approach claimed a performance gap within 0.2dB compared to the ML curve.

While prevailing neural decoding variants primarily focus on minimizing the average loss function, typically defined for the codeword bits, to reduce the resulting FER, limited effort has been directed towards adapting decoding failures to enhance their likelihood of being decodable in the post-processing of OSD. An exception is the hybrid BP-OSD framework \cite{jiang2007reliability}, which employs weighted extrinsic information to synthesize new bit reliability measurements for each decoding failure of BP, facilitating effective OSD post-processing. Similarly, focusing on the failed received sequences of BP, Zhang et al. \cite{zhang2023efficient} recently introduced an iterative quasi-decoder structure specifically tailored to generate new bit reliability measurements for successful OSD.

There is a consensus that OSD will eventually achieve ML decoding with a sufficiently large order $p$. The real challenge lies in maintaining superior performance given a constrained $p$, equivalent to limited computational resources in realistic applications. In line with the BP-OSD combination, our prior work \cite{li2023deep} addressed this challenge through the NMS-OSD framework. In this framework, a CNN model is employed, fed with the complete decoding trajectory of the NMS decoding failure, to synthesize a new measure. This measure aims to decrease the number of erroneous bits in the most reliable basis (MRB) or compact them into a smaller region of MRB. This process, known as decoding information aggregation (DIA), is followed by the presentation of an adaptive OSD which works by segmenting the MRB region to control the number of searching TEPs. Additional support comes from the application of an auxiliary criterion. This paper adopts the same hybrid framework to harness the strengths of both NMS and OSD. The main contributions of this study are outlined below:
\begin{itemize}
\item[*] Following a comprehensive design analysis of various NMS variants, considering both decoding FER and the impact of resulting decoding failures on subsequent OSD, it was revealed that all NMS variants can be seamlessly interpreted in the context of GNN. This is achieved by leveraging the permutation-invariant property and sharing technique. Importantly, they exhibit no distinguishable decoding performance, whether in the NMS phase or the OSD post-processing phase. Therefore, NMS-1 is chosen for its simplicity as the constituent for the hybrid NMS-OSD framework.

\item[*] Either of two alternatives is suggested as the crucial element of adaptive OSD about organizing TEPs effectively: a list of order patterns with a uniform size of TEPs , and a dynamic MRB segmenting scheme  based on the number of swapped bits during Gaussian elimination (GE) operations on the parity check matrix. For both methods, the execution priority of related order patterns is determined using a statistics-based rule extraction approach \cite{zhou2000statistics}.

\item[*] Considering the common requirement of larger maximum iterations for NMS decoding in longer LDPC codes, achieving iteration diversity gain for OSD is proposed. This is accomplished by partitioning the entire iteration trajectory of each NMS decoding failure into several groups and subsequently seeking out the optimal estimate across these groups.

\item[*] In a wide SNR region encompassing various short to moderate LDPC codes, extensive experimental simulations on the hybrid framework have solidified its merits, including low complexity, high throughput, and channel invariance. Furthermore, the proposed approach achieves decoding performance close to ML FER.
\end{itemize}

The rest of the paper is organized as follows. Section \ref{preliminary} presents necessary preliminaries about BP, NMS decoding, some OSD variants, and a key GNN property. Section \ref{motivation} elaborates on the motivations behind our work, while Section \ref{simulations} discusses the decoding performance and complexity analysis of selected codes. Finally, Section \ref{conclusions} concludes the paper with remarks and suggestions for further research.

%% file: sections/preliminary.tex
\section{Preliminaries}
\label{preliminary}

Assuming a binary message row vector $\mathbf{m} = [m_i]_1^K$ is given, it is encoded into a codeword $\mathbf{c} = [c_i]_1^N$ using $\mathbf{c} = \mathbf{mG}$ in the Galois field GF(2), where $K$ and $N$ are the lengths of the individual message and codeword, and the generator matrix $\mathbf{G}$ is commonly assumed to be full row rank.

Subsequently, a simple binary phase shift keying (BPSK) modulation scheme maps each bit $c_i$ to an antipodal symbol, given by $s_i = 1 - 2c_i$. Due to the additive white Gaussian noise (AWGN) $n_i$ with zero mean and variance $\sigma^2$, a corrupted sequence $\mathbf{y} = [y_i]_1^N$ is obtained, where $y_i = s_i + n_i$, and this becomes the channel output sent to the decoder for optimal estimation of the original codeword.

Considering the definition of the log-likelihood ratio (LLR) for the $i$-th bit as follows:
\begin{equation}
{l_i} = \log \left( \frac{{p(y_i|{c_i} = 0)}}{{p(y_i|{c_i} = 1)}} \right) = \frac{{2y_i}}{{\sigma^2}}
\end{equation}
it is evident that a larger magnitude of $y_i$ implies more confidence in the hard decision of the corresponding $i$-th bit. Notably, standard BP requires both $y_i$ and $\sigma^2$ for iterated LLR messages in the decoding phase, while some MS variants have the unique advantage of only requiring $y_i$, known as channel invariance \cite{lugosch18-1xVPf}.
\subsection{BP, MS Variants, and Their Neural Versions}
The bipartite Tanner graph of a code, closely related to its parity check matrix $\mathbf{H}$, consists of $\mathit{N}$ variable nodes, $\mathit{M}=N-K$ check nodes without loss of generality, and all edges connecting variable node $j$ and check node $i$ are indicative of any nonzero entry at row $i$ and column $j$ of $\mathbf{H}$.

For a specific LDPC code in application, standard BP, among many others, is a competitive decoder by exchanging directed decoding messages along edges of its Tanner graph, although the unavoidable short cycles resulting from the finite block length of the code degrade it into a sub-optimal decoder.

Supposing a plain flooding schedule for BP with at most $T$ iterations of decoding, then at the $t$-th iteration, $t \in \{1,\cdots T\}$, the message from variable node $v_i$ to check node $c_j$ is given by:
\begin{equation}
x_{v_i \to c_j}^{(t)}  = {l_i} + \sum\limits_{\substack{c_p \to v_i\\p \in \mathcal{C}(i)\backslash j}} {x_{c_p \to v_i}^{(t - 1)}}
\label{eq_v2c}
\end{equation}
while the reverse message from $c_j$ to $v_i$ is expressed as:
\begin{equation}
x_{c_j \to v_i}^{(t)} = 2{\tanh ^{ - 1}}\left( {\prod\limits_{\substack{v_q \to c_j\\q \in \mathcal{V}(j)\backslash i}} {\tanh \left( {\frac{{x_{v_q \to c_j}^{(t)}}}{2}} \right)} } \right)
\label{eq_c2v}
\end{equation}
where $\mathcal{C}(i)\backslash j$ denotes all neighboring check nodes of $v_i$ except $c_j$, and $\mathcal{V}(j)\backslash i$ denotes all neighboring variable nodes of $c_j$ except $v_i$, and all $x_{c_p \to v_i}^{(0)}$ terms in \ref{eq_v2c} are initialized to zero.

Then the alternation of \ref{eq_v2c} and \ref{eq_c2v} for all edges of the Tanner graph lasts in parallel $T$ times to fulfill the entire decoding process. Meanwhile, a posteriori value of the $i$-th bit after the $t$-th iteration is calculated as follows:
\begin{equation}
x_{v_i}^{(t)} = {l_i} + \sum\limits_{\substack{c_p \to v_i\\p \in \mathcal{C}(i)}} {x_{{c_p} \to {v_i}}^{(t - 1)}}
\label{eq_bit_decision}
\end{equation}
with the tentative hard decision of \ref{eq_bit_decision} yielding a tentative codeword estimation given by:
\begin{equation}
\widehat{\mathbf{c}}^{(t)}_{i}=\begin{cases}
 0& \text{ if } sgn(x_{v_i}^{(t)})= 1 \\ 
 1& \text{ otherwise } 
\end{cases}   
\end{equation}
where $sgn(\cdot)$ denotes the sign function. To speed up decoding, the process will commonly exit immediately once the early-stopping criterion $\widehat{\mathbf{c}}^{(t)}\mathbf{H}^{'}=\mathbf{0}$ is met.

A much simplified approximation \eqref{eq_ms} in the MS family was implemented as a substitution for \eqref{eq_c2v} to save the intensive computation of $\tanh$ or $\tanh^{-1}$ functions:
\begin{equation}
x_{c_j \to v_i}^{(t)} = s_{c_j \to v_i}^{(t)} \cdot \phi_{c_j \to v_i}^{(t)}
\label{eq_ms}
\end{equation}
where:
\begin{equation}
s_{c_j \to v_i}^{(t)} = \prod\limits_{\substack{v_q \to c_j\\ q \in \mathcal{V}(j)/i}} sgn\left( x_{v_q \to c_j}^{(t)} \right)
\label{sign_f}
\end{equation}
and:
\begin{equation}
\phi_{c_j \to v_i}^{(t)} = \min_{\substack{v_q \to c_j\\ q \in \mathcal{V}(j)/i}} \left| x_{v_q \to c_j}^{(t)} \right|
\label{min_f}
\end{equation}

To compensate for the performance loss of the MS decoder, the normalized MS and offset MS decoders were suggested in the literature by imposing a multiplicative weight or additive offset to the min-term of \eqref{eq_ms}. The effectiveness of this approach is justified by the fact that the magnitude of the min-term consistently overestimates the corresponding term of \eqref{eq_c2v}, thus shrinking it will generate a better correction.

Conventional BP can be readily transformed into a trellis structure by unrolling each decoding iteration in sequence, with a trainable parameter posed on each trellis edge. It is thus regarded as a specific NBP. For the neural version of any MS variant, denoted as the NMS family, although most edges can be weighted similarly to NBP, a subtle difference has to be emphasized: the min-term is a statistical quantity, not affiliated with any fixed edge. Thus, in the strict sense, weighting on the min-term is not equivalent to parameter sharing or tying in the context of neural networks. Since the NMS family shares an overwhelming advantage over NBP in terms of computational complexity while providing comparable decoding performance, our discussion will focus mainly on it hereafter. From some perspectives, the original normalized MS can be treated as a special case of the NMS family, with a sole parameter to be trained. It is also a good starting point when opting for the bottom-up design methodology to identify the efficacy of each parameter of an NMS variant \cite{li2023bottom}.

In the training phase, fed with batches of training data, an NMS decoder evolves progressively after its parameters are updated via the designated stochastic gradient descent (SGD) optimizer for a predefined loss function. A well-trained neural decoder is expected to be competent in narrowing the performance gap between BP and MS.
\subsection{OSD and Its Adaptation}

In one view, all OSD variants can be categorized in terms of whether GE is applied to the related $\mathbf{G}$ or $\mathbf{H}$. Our discussion will dwell on the H-centered OSD variants considering the actual performance duality. Notably, in the adopted hybrid framework, the OSD, playing the role of a post-processor, is merely requested to deal with the decoding failures of the preceded NMS.

\begin{figure}[htbp]
	\centering
\centerline{\includegraphics[width=0.3\textwidth]{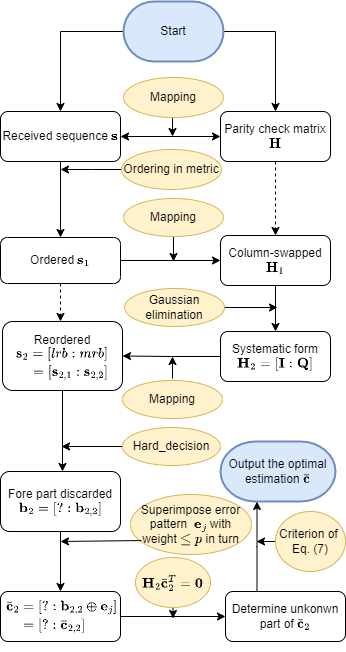}}
\caption{Diagram of OSD procedures}
\label{osd_diagram}
\end{figure}

For an order-$p$ conventional OSD, all TEPs with a Hamming weight of at most $p$ are allowed in the MRB part to be scrutinized later. The detailed procedures of OSD are diagrammed in Fig. \ref{osd_diagram}, where the ellipse items denote triggering events, solid and dotted lines imply proactive and passive resulting actions individually. Specifically, for a received sequence of soft information, all bits are bonded with columns of $\mathbf{H}$ first of all; then, they are sorted in ascending order of some reliability metric, resulting in the column swapping of $\mathbf{H}$ into $\mathbf{H_1}$ correspondingly. Next, the GE operation is solicited to reduce $\mathbf{H_1}$ into its systematic form $\mathbf{H_2}$ in GF(2). Apparently, the requested column swapping may abruptly disrupt the ascending order of involved bits locally.  Albeit, the existing sequence is partitioned into least reliable basis (LRB) and MRB parts. Now the hard-decision $\mathbf{b}_{2,2}$ of the MRB part is considered an anchoring point on which each TEP within the Hamming weight constraint is superimposed. Next, the vacant LRB part is derived from the parity check constraints. Lastly, all the forged $\mathbf{\bar{c}}_2$s compete for the optimal codeword estimate under the criterion of \eqref{argmin_dis}.

\begin{equation}
\label{argmin_dis}
\overline{\mathbf{c}} = \mathop {\arg\min } \sum\limits_{i = 1}^N {\mathbf{1}(\check{c}_i \ne {\hat{c}}_i)\left| y_i \right|}
\end{equation}
where $\mathbf{1}(\cdot)$ denotes the indicator function, $\mathbf{\check{c}}$ denotes each resulting candidate codeword after reversing all bit swaps of $\mathbf{\bar{c}}_2$ occurred in sorting or GE operations, and $\mathbf{\hat{c}}$ denotes the initial hard decision of the received sequence, respectively.

\begin{figure}[htbp]
	\centering
\centerline{\includegraphics[width=0.35\textwidth]{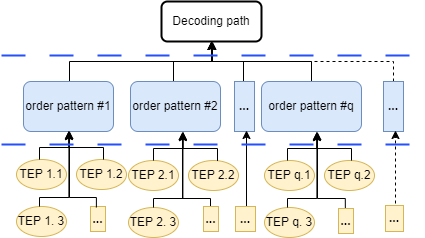}}
\caption{Hierarchical structure of decoding path}
\label{decoding_path_structure}
\end{figure}

Considering the list size of TEPs commonly dominates the OSD complexity, how to shrink it while retaining superior decoding performance is vital for OSD. In our prior work, we proposed an adaptive OSD to reorganize the TEPs, emphasizing the placement of more frequently occurring TEPs at the top priority list given limited computational resources. The enumerating of TEPs occurs in the form of a decoding path predetermined in advance, as shown in the hierarchical structure in Fig. \ref{decoding_path_structure}.
This decoding path consists of a list of dominant order patterns, each of which governs underlying TEPs of varying sizes. Notably, the precedence of order patterns and how to categorize the TEPs into individual order patterns significantly impact the final decoding performance.

Given the FER requirement, to lower the necessary Hamming weight of TEPs or concentrate to-be-flipped bits into the forefront of the MRB part, the Decoding Information Aggregation (DIA) technique was implemented via a CNN to synthesize a new reliability estimate for each codeword bit. An auxiliary criterion was proposed to further reduce the potential list size of TEPs. Interested readers can refer to \cite{li2023deep} for a detailed description.

%% file: sections/motivation.tex
\section{Motivations}
\label{motivation}

In the following, three LDPC codes will be discussed: CCSDS (128,64) code \cite{Book2012}, Wimax-like (384,192) code \cite{helmling19}, and WiMAX (802.16) (1056,880) code \cite{helmling19}. Regarding  the decoding failures of an NMS variant, the authentic error pattern refers to the corresponding discrepancy between the current hard-decision and ground-truth in terms of the MRB bits. Apparently, once the authentic error pattern is estimated correctly, the associated decoding failure is deemed decodable as a result.

\subsection{How to design and interpret NMS variants?}
For NBP or fully weighted NMS of LDPC codes, it often outperforms standard BP marginally. However, the associated cost involves numerous multiplications due to weighting every decoding message, making it less appealing as a decoder design solution. Given that the curse of dimensionality is typically encountered in the field of channel coding, it is almost impossible to determine a finite number of parameters that can cope well with any input sequence to the associated NN. In other words, when treating decoding as a label classification task, accurately drawing the discriminating boundary between the labels becomes intractable through training the NN, especially considering the exponential growth of labels with the increase in code length.

To this end, we adopt the mentioned bottom-up design on GNN. In other words, we no longer seek NMS solutions that significantly enhance decoding performance. Instead, our primary goal is to choose a low-complexity NMS variant that not only provides comparable decoding performance but also ensures that the resulting decoding failures have good odds within the scope of OSD.

It is noticed that GNNs emphasize preserving the permutation-invariant property for the related operations in most cases. Hence, it enlightens us to comply with this rule when designing any new NMS variant, which leads to augmenting new parameters as substitutions of \eqref{eq_v2c}, \eqref{eq_bit_decision}, and \eqref{eq_ms} as follows:

\begin{equation}
x_{v_i \to c_j}^{(t)}  = {\zeta_1} \cdot {l_i} + \sum\limits_{\substack{c_p \to v_i\\p \in {\mathcal{C}(i)/j}}} {x_{c_p \to v_i}^{(t - 1)}}
\label{eq_v2c_weight}
\end{equation}

\begin{equation}
x_{v_i}^{(t)} = \zeta_2 \cdot {l_i} + \sum\limits_{\substack{c_p \to v_i\\p \in \mathcal{C}(i)}} {x_{{c_p} \to {v_i}}^{(t - 1)}}
\label{eq_bit_decision_weight}
\end{equation}

\begin{equation}
x_{c_j \to v_i}^{(t)} = \zeta_3 \cdot s_{c_j \to v_i}^{(t)} \cdot  {\phi}_{c_j \to v_i}^{(t)}
\label{eq_ms_weight}
\end{equation}

In view of \eqref{eq_v2c_weight}, \eqref{eq_bit_decision_weight}, and \eqref{eq_ms_weight}, NMS-1, equivalently the original normalized MS in the literature, states that $\zeta_3$ is the sole trainable parameter with constant $\zeta_1 = \zeta_2 = 1$. NMS-2 actually includes two trainable parameters with $\zeta_{1} = \zeta_{2}$ and $\zeta_{3}$. NMS-3 accounts for the case of independent $\zeta_{1,2,3}$ parameters, and NMS-$r$, besides two other parameters $\zeta_{1}$ and $\zeta_{2}$, involves a small FCN implementation of two layers to substitute for the product of $\zeta_3 \cdot {\phi}_{c_j \to v_i}^{(t)}$ in \eqref{eq_ms_weight}, where $r$ points out that the number of sorted magnitudes as its input is the cardinality of the set $\mathcal{V}(j)/i$.

\input{plots/fer_128_NMS_options}

In the case of the (128,64) code, the FER curves of well-trained NMS variants in Fig.\ref{fer_128_options} revealed two key observations:

\begin{itemize}
  \item Generally, all NMS variants lag behind BP in terms of FER performance. However,  NMS variants exhibit nearly identical decoding performance.
  \item The postprocessing performance of OSDs for decoding failures left by individual NMS variants is also similar, given identical external settings such as the support of DIA, auxiliary criterion application, and iteration diversity.
\end{itemize}

Similar observations were made in further experiments conducted on (384,192) code and (1056,880) code. Table \ref{tab:para_settings} presents the parameter evaluations for several NMS variants after full training on the three codes. It is found that $\zeta_3$ is the only parameter that makes a significant difference, while $\zeta_{1,2}$ can be roughly set to one for simplicity, as the NMS variants are not sensitive to small changes in these parameters. Consequently, NMS-1 is favored for its simplicity to be chosen as the default component in the hybrid NMS-OSD framework.

\subsection{Two Alternatives of Adaptive OSD}
As illustrated in Fig. \ref{decoding_path_structure}, the decoding path consists of a list of order patterns, each governing a number of TEPs.

Decoding failures are not equally probable, as indicated by the occurrences of associated authentic error patterns covered in the order patterns during the validation phase. Considering the constraint on computational complexity, there is a need to truncate the list of order patterns. It is essential to determine how to allocate TEPs into each order pattern and establish priorities between order patterns beforehand. This ensures that the TEPs in the remaining order patterns can cover potential authentic error patterns in the test phase with the maximum average probability.

Two schemes are devised for populating the order patterns with partitioned TEPs. One generates order patterns of roughly uniform size, while the other creates a list of order patterns with varying counts of TEPs. We rely on empirical results to identify the priority of each order pattern. For both schemes, the categorization of TEPs into each order pattern is based roughly on the Hamming weight they possess, despite some different technical details as described below. Specifically, in a typical waterfall Signal-to-Noise Ratio (SNR) region for a specific code, with available statistics of authentic error patterns after sufficient sampling, we intuitively assign the highest priority to order patterns covering the most authentic error patterns in total and so forth. This prioritization is necessary when limited computational resources mandate the suppression of the number of searching TEPs.

\subsubsection{Evenly Partitioning TEPs}
Suppose the Hamming weight limit being \(p\) for each TEP, then we can partition all qualified TEPs into blocks, or termed order patterns for consistency in the context, with each order pattern of approximately uniform size \(w_b\).

For instance, in the case of a (128,64) code, suppose \(w_b=32\) with \(p=3\), then the order patterns may be populated as follows: the first order pattern includes only one all-zero TEP exceptionally; then all TEPs of Hamming-weight 1 are confluent into two order patterns sequentially, and so forth. The general categorizing rule is that TEPs with the nearest indices sum of the non-zero elements in between are gathered in the same order pattern unless it is full. Lastly, the prioritizing of order patterns is identified empirically. Notably, at the extreme, we can set \(w_b=1\) to denote one TEP being one order pattern, conceptually implying the finest granularity of scheduling for TEPs partition.

Compared with the following dynamic scheme, this method is much simpler. However, the drawback is that with the increase of the code block length, accurately evaluating the priority of each order pattern rapidly becomes demanding. For instance, in the case of a (1056,880) code, the number of its Hamming-weight 3 TEPs exceeds 100 million, resulting in more than one million order patterns given \(w_b=32\), hence a potentially heavy load in pinpointing the priorities of order patterns empirically. Actually, in the mentioned scenario, it is formidable to list all the TEPs since it commonly crashes computer memory capacity, not to mention the manipulations on them. Additionally, due to the dispersed distribution of authentic error patterns for longer codes, a large volume of decoding failures is requested to discriminate the priorities of associated order patterns via manifesting the statistical differences between them.

Therefore, accurately querying the priorities of order patterns is a time-consuming task in real scenarios, despite its one-shot feature. We opt for this method for short (128,64) and (384,192) codes while leaving the longer (1056,880) code for the next alternative.

\subsubsection{Dynamically Partitioning TEPs}

Partitioning an MRB into smaller units is a crucial step in constructing this OSD adaptation, as the list of order patterns that accommodate TEPs is closely associated with the partition. However, determining a fixed partition for the MRB region of a specified code that adapts well to any decoding failure is commonly challenging.

\begin{figure}[htbp]
\centering
\centerline{\includegraphics[width=0.35\textwidth]{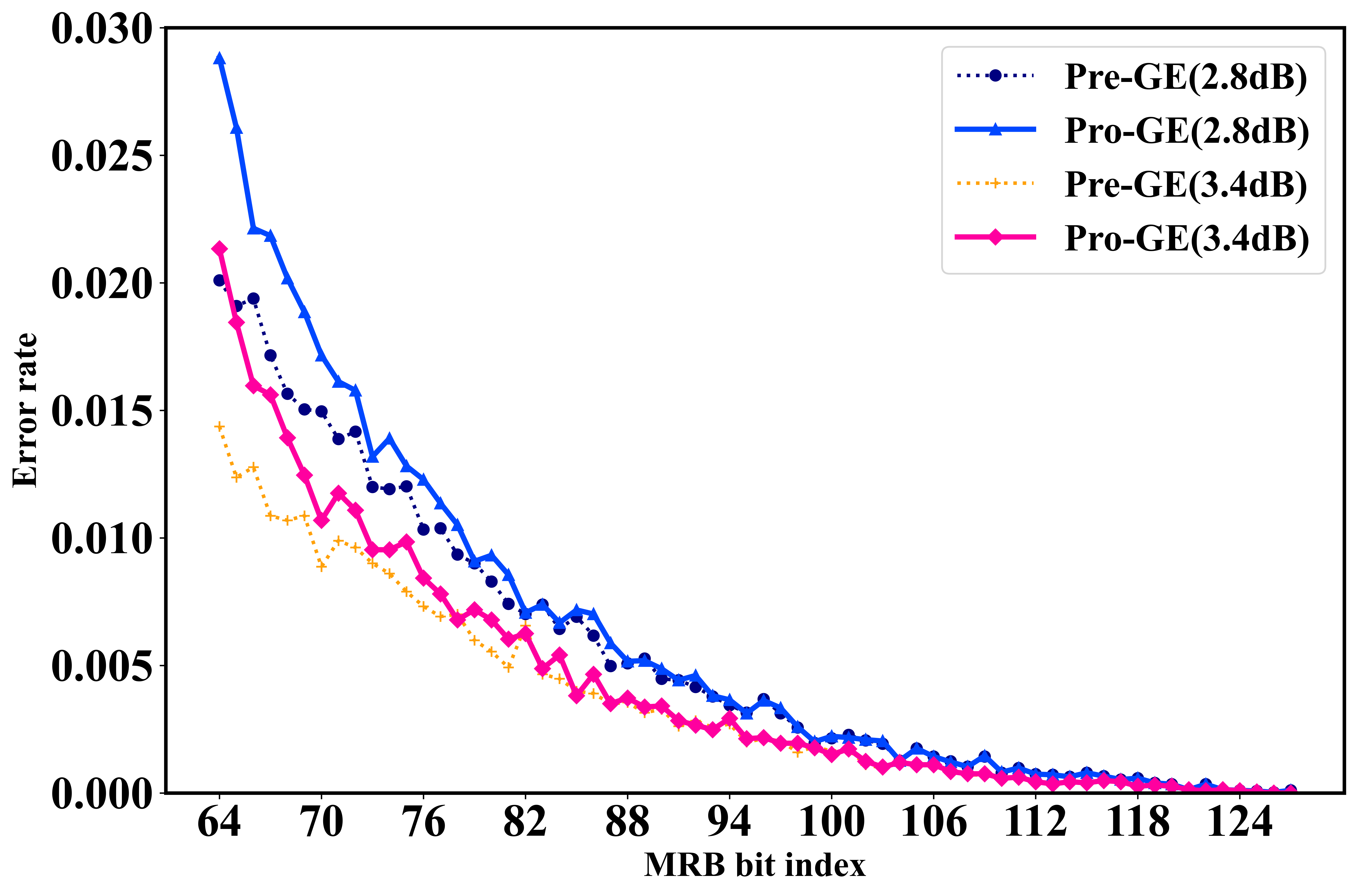}}
\caption{Statistics about the error rates of all MRB bits before and after GE calling are compared for LDPC (128,64) code at SNR=2.8, 3.4dB. Suppose the starting MRB index is 64; after some oscillation, it is found that the diverging curves will re-converge at some index 88.}
\label{swap_128}
\end{figure}

On one hand, the first step of OSD is to reduce \(\mathbf{H}\) into its systematic form via binary GE operation, entailing the swapping of bits between the LRB and MRB parts. Apparently, the swapping raises the risk of exceeding the OSD capability, as some unreliable bits originally located in the LRB are exchanged into the MRB. For instance, in the case of a (128,64) code at SNR=2.8, 3.4dB, for the first 24 sorted MRB bits in ascending order of magnitude, the bit error rate before GE operation and after GE operation was compared in Fig.~\ref{swap_128}; it shows the swapping due to GE operation leads to the hazard of more erroneous hard decisions in MRB, which may severely undermine OSD decoding that allows for Hamming weight at most \(p\) searching TEPs.

On the other hand, let \(\rho_s\) denote the number of swapped columns (also swapped bits) as a random variable, and its mean and standard deviation are presented empirically in Table~\ref{comp_swap_dist}, revealing that more swapped columns are commonly requested with the increase of \(\mathbf{H}\) dimensions. Furthermore, \(\rho_s\) is not sensitive to the alteration of SNR points once the code is specified. The last column of the table shows the maximum number of affected MRB bits after sorting for LDPC (128,64) code and (1056,880) code, indicating that only a small proportion of MRB bits are affected by the GE operation.
\begin{table}
\centering
\caption{Statistics comparison of the number of swapped bits in reducing  $\mathbf{H}$ into its systematic form via GE operation for LDPC (128,64) and (1056,880) codes \cite{helmling19}}
\begin{tabular}{|c|c|c|c|c|}
\hline
Code                        & SNR   & Mean & Std  & \begin{tabular}[c]{@{}c@{}}Affected farthest \\ MRB index (since zero)\end{tabular} \\ \hline
\multirow{2}{*}{(128,64)}   & 2.8dB & 2.48 & 1.37 & 12          \\ \cline{2-5} 
                            & 3.4dB & 3.43 & 1.66 & 12          \\ \hline
\multirow{2}{*}{(1056,880)} & 3.2dB & 35.1 & 6.04 & 44          \\ \cline{2-5} 
                            & 3.6dB & 38.2 & 6.14 & 40          \\ \hline
\end{tabular}
\label{comp_swap_dist}
\end{table}

Concerning the swapped-in MRB bits of decoding failures, their higher error probability and varied counts prompt us to dynamically set the boundary of indices of non-zero elements of TEPs. Specifically, we denote the range of the indices of non-zero elements of any qualified TEPs in the form of $[0, d_1, d_2, d_3)$, where $d_1$ and $d_2$ are intermediate delimiting points with $d_1 = d_0 + \min \{\rho_s, 5\}$ and $d_2 = d_1 + \min \{2\rho_s, 10\}$, and $d_0$ is suggested in the last column of Table~\ref{comp_swap_dist}. The right endpoint index $d_3$ is supposed to be determined empirically in the range of $[d_2, K-1]$. Next, to constrain complexity, we assign appropriate Hamming weight limits $[\xi_{m1}, \xi_{m2}, \xi_{m3}]$, alongside the constraint $\sum_{i=1}^{3}\xi_{mi}\le p$, for the triple intervals $[0, d_1), [d_1, d_2), [d_2, d_3)$, respectively.

At this point, all legitimate order patterns $[\xi_{1}, \xi_{2}, \xi_{3}]$ under the constraints $\xi_{j} \le \xi_{mj}, j=1,2,3$ can be uniquely identified. Consequently, each $\xi_j$ regulates how many non-zero elements are allowed, whose indices are confined to the designated interval. Lastly, we populate the chosen order patterns with qualified TEPs. Depending on the $\rho_s$ evaluation on the fly, the included TEPs are customized for each specific decoding failure. Additionally, for the sake of curbing complexity, the evaluation of $d_3$ may be subject to the evaluation of $\xi_3$. For example, $d_3=880$ if $\xi_3=1$ while $d_3=200$ if $\xi_3=3$ for a (1056,880) code. Afterward, these order patterns are prioritized empirically by sampling sufficient decoding failures to probe the statistics of the authentic error patterns covered by the corresponding order patterns.

As indicated in Table~\ref{pattern_dist} for a (128,64) code at SNR=2.8dB, the decoding path of each decoding failure varies with $\rho_s$ learned in GE operation. Then suppose the nominal length of the decoding path is 4, or equivalently, at most 4 order patterns are taken into account, besides the constraints $\xi_{m1}=2, \xi_{m2}=1, \xi_{m3}=1$, then the OSD decoding follows the path of $[0,0,0]\to[1,0,0]\to [0,1,0]\to[1,1,0]$ for the $\rho_s=3$ cases, while $[0,0,0]\to [1,0,0]\to[0,1,0]\to[2,0,0]$ for the $\rho_s=4$ cases. Only 4 types of order patterns were observed for the $\rho=11$ cases due to its rareness. Meanwhile, any order pattern unqualified with the constraints will be skipped in visiting the decoding path. In this way, we can conveniently balance the trade-off between performance and complexity by tweaking the parameters such as $p$, $\xi_{mi}$, $d_3$, or the nominal length of the decoding path.

\begin{table}
\centering
\caption{The order patterns, identified as the components of a decoding path, are sorted by the statistical occurrences of categorized authentic error patterns after GE operation for LDPC (128,64) code at SNR=2.8dB} 

\begin{tabular}{|c|clll|}
\hline
\begin{tabular}[c]{@{}c@{}}\# of \\ swaps\end{tabular} &
  \multicolumn{4}{c|}{First 6 leading  order patterns} \\ \hline
0 & \multicolumn{4}{c|}{\cellcolor[HTML]{FFFFFF}{[}0, 0, 0{]}, {[}0, 1, 0{]}, {[}1, 0, 0{]}, {[}1, 1, 0{]}, {[}0, 0, 1{]}, {[}2, 0, 0{]}} \\ \hline
1 &
  \multicolumn{4}{c|}{{[}{[}0, 0, 0{]}, {[}1, 0, 0{]}, {[}0, 1, 0{]}, {[}1, 1, 0{]}, {[}0, 0, 1{]}, {[}2, 0, 0{]}{]}} \\ \hline
2 &
  \multicolumn{4}{c|}{{[}{[}0, 0, 0{]}, {[}1, 0, 0{]}, {[}0, 1, 0{]}, {[}1, 1, 0{]}, {[}2, 0, 0{]}, {[}0, 0, 1{]}{]}} \\ \hline
3 &
  \multicolumn{4}{c|}{{[}{[}0, 0, 0{]}, {[}1, 0, 0{]}, {[}0, 1, 0{]}, {[}1, 1, 0{]}, {[}2, 0, 0{]}, {[}0, 0, 1{]}{]}} \\ \hline
4 &
  \multicolumn{4}{c|}{{[}{[}0, 0, 0{]}, {[}1, 0, 0{]}, {[}0, 1, 0{]}, {[}2, 0, 0{]}, {[}1, 1, 0{]}, {[}0, 2, 0{]}{]}} \\ \hline
5 &
  \multicolumn{4}{c|}{{[}{[}0, 0, 0{]}, {[}1, 0, 0{]}, {[}0, 1, 0{]}, {[}2, 0, 0{]}, {[}1, 1, 0{]}, {[}0, 0, 1{]}{]}} \\ \hline
... &
  \multicolumn{4}{c|}{...} \\ \hline
11 &
  \multicolumn{4}{c|}{{[}{[}0, 0, 0{]}, {[}1, 0, 0{]}, {[}0, 1, 0{]}, {[}0, 0, 1{]}{]}} \\ \hline
\end{tabular}
\label{pattern_dist}
\end{table}
\subsection{Iteration Diversity for Longer Codes}

To harness the full decoding power of the NMS family, the maximum number of iterations $T$ setting for an NMS variant typically increases with the code block length. This observation motivates us to split the decoding iteration trajectory into smaller groups and use them to train separate DIA models. These smaller models are more manageable in size compared to a single, larger DIA model without splitting.
\begin{figure}[htbp]
	\centering
	\centerline{\includegraphics[width=0.35\textwidth]{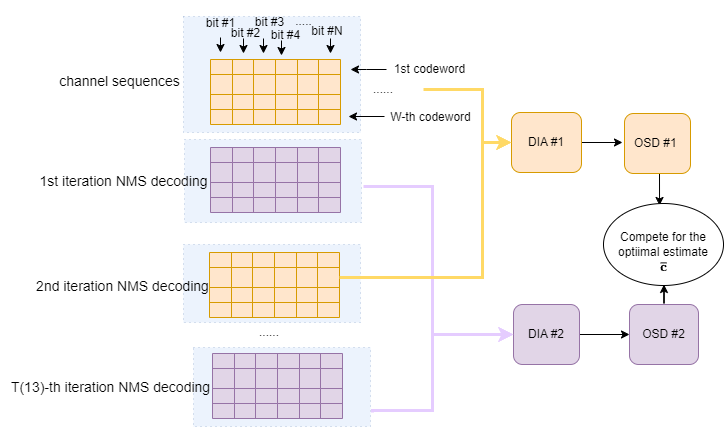}}
	\caption {Block diagram of the approach to achieve diversity gain through iteration splitting for training DIA models. In this illustration, the decoding trajectory of length $T=13$ is split into 2 groups per decoding failure.}
	\label{diversity_iteration}
\end{figure}
As illustrated in Fig.~\ref{diversity_iteration}, when $T=13$, a possible bi-group solution could be $g_e=\{0,2,4,\ldots,12\}$ and $g_o=\{1,3,5,\ldots,13\}$, where $g_e$ and $g_o$ represent iteration tokens conveying information about posteriors of all codeword bits at the indicated iterations. This arrangement is equivalent to partitioning a long time series into two shorter ones. Therefore, diversity gain can be achieved by employing several small DIA models with these shorter time series. The optimal estimate is then selected from the merged estimates of individual OSDs fed with the outputs of individual DIA models.

%% file: plots/fer_128_NMS_options.tex
\begin{figure}[htbp]
\centering
\begin{tikzpicture}
\begin{semilogyaxis}[
    scale = 0.75,
    xlabel={$E_b/N_0$(dB)},
    ylabel={FER},
    xmin=2.2, xmax=3.3,
    ymin=3e-4, ymax=5e-1,
    xtick={2.2,2.2,2.4,2.6,...,3.2},
    legend pos = south west,
    ymajorgrids=true,
    xmajorgrids=true,
    grid style=dashed,
    legend columns=2,
    legend style={font=\tiny},
    ]
%1 plot for BP (FER)
\addplot[
color=blue,
mark= +,
dashed,
]
coordinates {
(2.5,0.13)
(2.75,0.085)
(3.0,0.053)
(3.25,0.027)
(3.5,0.013)
(3.75,0.005)
(4.0,0.002)
};	
\addlegendentry{BP(T=40)}

%0 plot for NMS-1 (FER)
\addplot[
color=violet,
mark=x,
dashed,
very thin
]
coordinates {
(2.2,0.3445)
(2.4,0.2666)
(2.6,0.1968)
(2.8,0.1390)
(3.0,0.0948)
(3.2,0.0593)
};	
\addlegendentry{NMS-1(T=13)}

%0 plot for NMS-2 (FER)
\addplot[
color=red,
mark=halfcircle,
dashed,
very thin
]
coordinates {
(2.2,0.3451)
(2.4,0.2672)
(2.6,0.1973)
(2.8,0.1394)
(3.0,0.0950)
(3.2,0.0594)
};	
\addlegendentry{NMS-2(T=13)}

%0 plot for NMS-3 (FER)
\addplot[
color=orange,
mark=diamond,
dashed,
very thin
]
coordinates {
(2.2,0.3422)
(2.4,0.2644)
(2.6,0.1951)
(2.8,0.1379)
(3.0,0.0938)
(3.2,0.0584)
};	
\addlegendentry{NMS-3(T=13)}

%0 plot for NMS-r (FER)
\addplot[
color=cyan,
mark=triangle,
dashed,
very thin
]
coordinates {
(2.2,0.3509)
(2.4,0.2743)
(2.6,0.2018)
(2.8,0.1452)
(3.0,0.0991)
(3.2,0.0625)
};	
\addlegendentry{NMS-r(T=13)}
%OSD-order3-Groups100-size32-DIA-Auxilary-diversify2	for NMS-1 failures
\addplot[
color=violet,
mark=asterisk,
very thin
]
coordinates {
(2.2,0.02349)
(2.4,0.01648)
(2.6, 0.01065) 
(2.8, 0.00809) 
(3.0, 0.00518) 
(3.2, 0.00288) 
};	
\addlegendentry{OSD(NMS-1)}

%OSD-order3-Groups100-size32-DIA-Auxilary-diversify2	for NMS-2 failures
\addplot[
color=red,
mark=halfcircle*,
very thin
]
coordinates {
(2.2,0.02222)
(2.4,0.01566)
(2.6, 0.01063) 
(2.8, 0.00807) 
(3.0, 0.00566) 
(3.2, 0.00297) 
};	
\addlegendentry{OSD(NMS-2)}
%OSD-order3-Groups100-size32-DIA-Auxilary-diversify2	for NMS-3 failures
\addplot[
color=orange,
mark=diamond*,
very thin
]
coordinates {
(2.2,0.0222)
(2.4,0.01652)
(2.6, 0.01124) 
(2.8, 0.00764) 
(3.0, 0.00523) 
(3.2, 0.00328) 
};	
\addlegendentry{OSD(NMS-3)}

%OSD-order3-Groups100-size32-DIA-Auxilary-diversify2	for NMS-r failures
\addplot[
color=cyan,
mark=triangle*,
very thin
]
coordinates {
(2.2,0.02128)
(2.4,0.01583)
(2.6, 0.0105) 
(2.8, 0.00748) 
(3.0, 0.00494) 
(3.2, 0.00296) 
};	
\addlegendentry{OSD(NMS-r)}

\end{semilogyaxis}
\end{tikzpicture}
\caption{FER comparison for various NMS variants and corresponding OSD with support of DIA, auxiliary criterion and iteration diversity for LDPC  (128,64) code}
\label{fer_128_options}
\end{figure}

%% file: sections/simulations.tex
\section{Simulation Results and Complexity Analysis}
\label{simulations}

For all LDPC codes (128,64), (384,192), and (1056,880), we employ the hybrid NMS-OSD framework. The NMS component can be either NMS-1 or NMS-2, and one OSD adaptation utilizes uniform-sized order patterns for the first two codes, while the strategy of dynamically partitioned TEPs is applied for order patterns for the last code. All simulations are executed on the Tensorflow or Colab online platform provided by Google. The related source code will be made available on the GitHub website after filing.

\subsection{Training Logistics for NMS Model and DIA Model}
For the training of the NMS model, we use the accumulated sum of cross-entropy from the literature \cite{nachmani18} as the loss definition, given by
\begin{equation}
\resizebox{.8\hsize}{!}{$
\begin{aligned}
\ell (\mathbf{c},\bar{\mathbf{c}}) = \frac{1}{T}\sum\limits_{j=1}^{T}\sum\limits_{i=1}^{N} \sum\limits_{z=0}^1 \left(p(c_i = z)\cdot\log \frac{1}{{p(\bar{c}_i^{(j)} = z)}} \right)	
\end{aligned}
$}
 \nonumber
\end{equation}
where $\mathbf{c}$ and $\bar{\mathbf{c}}$ denote the ground-truth and the optimal codeword estimate. And the general cross-entropy definition for DIA model training.

For the SGD implementation, we use the conventional Adam optimizer \cite{kingma14} to reduce the loss function by updating the involved parameters in mini-batch mode. Due to the sparseness of parameters for NMS variants, the parameter optimization is not sensitive to the learning rate. Hence, the initial value of $0.01$ with a decaying factor of $0.95$ every $500$ training steps is adopted. A batch of training data is generated by a blend of samples from the typical SNR waterfall region \cite{li2023bottom}. For simplicity, the trained parameters will be deployed universally for all SNR points in the test phase. As for the DIA model without iteration diversity, it is a four-layered simple CNN model \cite{li2023deep} trained by feeding the decoding failure trajectories of the NMS variant. Once iteration diversity is preferred, each resulting small DIA model is implemented via a simple two-layered CNN model with a narrower width.
\begin{table}[htbp]
\caption{Trained parameters or settings initialized with constant ones for various codes in the NMS-OSD architecture}
\label{tab:para_settings}
\resizebox{\linewidth}{!}{
\begin{tabular}{|c|c|c|c|c|c|c|c|c|}
\hline
Codes      & \begin{tabular}[c]{@{}c@{}}NMS \\ variants\end{tabular} & \begin{tabular}[c]{@{}c@{}}Training  \\ SNR range\end{tabular} & Batch size & Iterations & $\zeta_1$ & $\zeta_2$ & $\zeta_3$ & \begin{tabular}[c]{@{}c@{}}\# of  \\ diverse groups\end{tabular} \\ \hline
(128,64)   & NMS-1                                                   & {[}2.2dB,3.2dB{]}                                             & 100        & 13         & 1.        & 1.        & 0.644     & 2                                                               \\ \hline
(384,192)  & NMS-2                                                   & {[}1.5dB,2.5dB{]}                                             & 50         & 20         & 0.98      & 0.98      & 0.77      & 3                                                               \\ \hline
(1056,880) & NMS-2                                                   & {[}3.0dB,3.8dB{]}                                             & 35         & 16         & 0.96      & 0.96      & 0.72      & -                                                               \\ \hline
\end{tabular}
}
\end{table}

\begin{table}[htbp]
\caption{Settings for various OSD schemes and observed number of searched TEPs on average for three LDPC codes}
\label{tab:osd_setting}
\resizebox{\linewidth}{!}{
\begin{tabular}{|c|c|c|c|}
\hline
Codes                       & OSD schemes                                                                                  & \begin{tabular}[c]{@{}c@{}}\# of \\ Order patterns\end{tabular} & \begin{tabular}[c]{@{}c@{}}TEP list size\\ in SNR regions \\ under test\end{tabular} \\ \hline
\multirow{3}{*}{(128,64)}   & \multirow{3}{*}{\begin{tabular}[c]{@{}c@{}}Order patterns \\ of uniformed TEPs\end{tabular}} & 30                                                              & 899.7-986.2                                                                          \\ \cline{3-4} 
                            &                                                                                              & 100                                                             & 2826.2-3096.0                                                                        \\ \cline{3-4} 
                            &                                                                                              & 200                                                             & 5504.8-6026.5                                                                        \\ \hline
\multirow{4}{*}{(384,192)}  & Conventional order-3 OSD                                                                     & -                                                               & 1,179,809                                                                            \\ \cline{2-4} 
                            & \multirow{3}{*}{\begin{tabular}[c]{@{}c@{}}Order patterns \\ of uniformed TEPs\end{tabular}} & 80                                                              & 2575.3-4249.6                                                                        \\ \cline{3-4} 
                            &                                                                                              & 200                                                             & 5468.9-9418.5                                                                        \\ \cline{3-4} 
                            &                                                                                              & 300                                                             & 7738.4-13451.8                                                                       \\ \hline
\multirow{2}{*}{(1056,880)} & Conventional order-3 OSD                                                                     & -                                                               & 113,579,401                                                                          \\ \cline{2-4} 
                            & \begin{tabular}[c]{@{}c@{}}Order patterns \\ of dynamically \\ allocated TEPs\end{tabular}   & 12                                                              & around 223k                                                                          \\ \hline
\end{tabular}
}
\end{table}
Some related parameter settings of NMS variants and OSD settings are listed in Table.\ref{tab:para_settings} and Table.\ref{tab:osd_setting} respectively for reference.
\subsection{Decoding Performance}

In the legends of the following figures, NMS-OSD($G_r$=xx) denotes the combination of default NMS-1 with iterations displayed in Table.\ref{tab:para_settings} and OSD of a decoding path consisting of xx order patterns, supported by the DIA and the auxiliary criterion. Notably, it is validated that marginal improvement is observed only when the iterations setting T increases to 40 for the NMS decoding alone of the discussed codes.
\input{plots/fer_128_NMS_diversify}

For the (128,64) code, the decoding performance of various schemes across the waterfall region is presented in Fig.\ref{fer_128_diversify} with the universal parameters trained in the range of $2.2-3.2$ dB. It is found that the FER curve of BP with $T=40$ lags far behind NBP-D(10,4,4) \cite{buchberger21}. Yet, the latter is surpassed by another 0.4dB by the scheme NMS-OSD($G_r$=30) at FER=$10^{-3}$, which, in turn, lags behind by about 0.2dB compared with $G_r$=100 of its kind. When $G_r$ further increases to 200, its performance approaches the SOTA decoder $D_{10}$-OSD-2 \cite{Rosseel2022}, which is within a 0.2dB gap of ML decoding \cite{helmling19}. 
\input{plots/fer_384_NMS_diversify}

Compared with the (128,64) code, the prolonged code length enables the waterfall region of the (384,192) code of the same rate to lower to 1.5-2.5dB, implying more tolerance to harsher channel conditions. As shown in Fig.\ref{fer_384_diversify}, NMS alone lags behind BP by about 0.2dB, but its combination with OSD of $G_r=80$ reverses the situation by leading by more than 0.4dB over the latter at FER=$10^{-3}$. Further expanding $G_r$ can push the improvement slowly towards the ML curve, as indicated by the attempts of $G_r=200,300$. To narrow the existing performance gap of 0.4dB between our method with ML, it is expected that a substantially large $G_r$ is required to fulfill the target, which, however, implies a heavy computational load in terms of the number of searching TEPs. As a side note, at a much expensive cost, the hybrid combination of the NMS and conventional order-3 OSD without DIA assistance manifested inferior performance by about 0.1dB than our NMS-OSD($G_r=80$), which partly signifies the necessity of applying DIA.

\input{plots/fer_1056_NMS_dynamics}

Compared with the above two codes, the (1056,880) code has a much longer block length as well as a higher rate of 0.83. Apparently, its OSD decoding becomes more challenging for the reason of the significantly expanded searching space and dispersed authentic error patterns. To alleviate the complexity burden, the dynamical searching TEPs mechanism is preferably applied for the chosen OSD. As shown in Fig.\ref{fer_1056}, it is observed that the NMS variant alone lags behind BP by about 0.1dB, while its combination with OSD leads the latter by about 0.2dB without DIA (signified by the suffix -N in the legend). In comparison, another 0.1dB gain at FER=$10^{-3}$ is observed with the aid of the DIA model. Haunted by the potential surging computational complexity, the constraints $\xi_{m1}\leq3$, $\xi_{m2}\leq2$, $\xi_{m3}\le2$, and $\sum_{i=1}^{3}\xi_{mi}\le3$ are imposed on this dynamic OSD version. Consequently, all designed schemes lag behind ML by beyond 0.25dB in our experimental simulations.

\subsection{Complexity Analysis}

With respect to OSD complexity, though extensive binary operations such as 'xor' or 'and' are called for in GE operations, we deliberately ignore them for their hardware-friendly nature, along with other sorting complexities scattered throughout the OSD procedures. Instead, we focus on managing the size of the searching TEPs, each of which is accompanied by several real number additions in seeking the optimal estimate according to Equation \ref{min_f}, accounting for the bulk of OSD complexity.

As listed in the last column of Table.\ref{tab:osd_setting}, the proposed fixed-TEP-size or dynamic-TEP-size OSD versions can effectively reduce the list size of candidate TEPs compared with their conventional counterparts. Specifically, with support of DIA, auxiliary criterion, and iteration diversify, throughout the waterfall region [2.2dB,3.2dB] of (128,64) code, the requested number of searching TEPs ranges from 5504 to 6026 for the fixed-TEP-size OSD of nominal $G_r=200$ per NMS decoding failure, while it ranges from 899 to 986 for the case of $G_r=30$. 
As a comparison, a conventional order-3 OSD requests 43745 searching TEPs. For (384,192) code, the average number of requested searching TEPs fluctuates between 7738-13451 in its waterfall region, while the figure surges to more than 1 million for the conventional order-3 OSD, not to mention the latter's inferior decoding performance as illustrated in Fig.\ref{fer_384_diversify}.

For the longer (1056,880) code, the computational saving for the dynamic-TEP-size OSD over the conventional one grows enormously as well. As indicated in the same table, with the support of DIA, the number of requested searching TEPs for the proposed framework is about two orders of magnitude less than its conventional counterpart, besides a leading 0.1dB decoding advantage for the former.

Whether with iteration diversity or not, the requested DIA models are simple 2-layered or 4-layered CNN models, each of which accounts for trainable parameters of fewer than two hundred. In contrast, the SOTA decoder of $D_{10}$-OSD-2(25) \cite{Rosseel2022} requests a bunch of BP-RNN decoders to be trained, whose trainable parameters sum up to $10240$. Therefore, our approach can greatly relieve the training load, besides the significant advantage in terms of implementation complexity. On the other hand, in the test phase, the SOTA decoder requires BP and OSD to interact between them several times per iterative decoding, which may severely impair the framework throughput. In comparison, our proposed method leaves the bulk of the decoding load on the fast NMS variant and only appeals for the support of the OSD with DIA for the small amount of NMS decoding failures, thus facilitating high throughput significantly.

%% file: plots/fer_128_NMS_diversify.tex
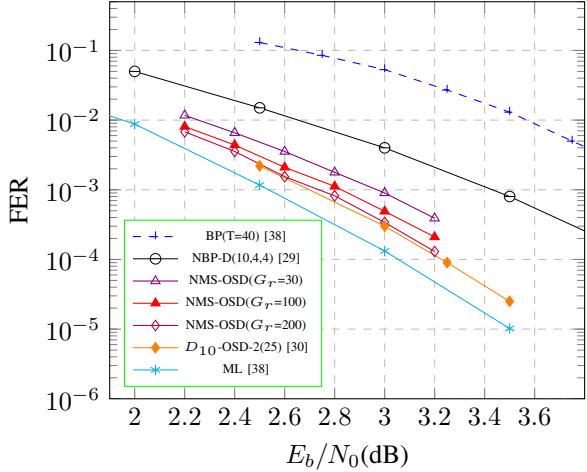
\begin{figure}[htbp]
\centering
\begin{tikzpicture}
\begin{semilogyaxis}[
    scale = 0.75,
    xlabel={$E_b/N_0$(dB)},
    ylabel={FER},
    xmin=1.9, xmax=3.8,
    ymin=1e-6, ymax=5e-1,
    xtick={2.0,2.2,2.4,2.6,...,3.8},
    legend pos = south west,
    ymajorgrids=true,
    xmajorgrids=true,
    grid style=dashed,
    legend style={font=\tiny},
    ]
%1 plot for BP (FER)
\addplot[
color=blue,
mark= +,
dashed,
]
coordinates {
(2.5,0.13)
(2.75,0.085)
(3.0,0.053)
(3.25,0.027)
(3.5,0.013)
(3.75,0.005)
(4.0,0.002)
};	
\addlegendentry{BP(T=40) \cite{helmling19} }
%3 plot for NBP-D(10,4,4) FER	
\addplot[
color=black,
mark=halfcircle,
very thin
]
coordinates {
(2.0,5e-2)
(2.5,1.5e-2)
(3.0,4e-3)
(3.5,8e-4)
(4.0,1.2e-4)
};
\addlegendentry{ NBP-D(10,4,4) \cite{buchberger21}}
%plot for NMS-1,T=13, OSD of order-pattern size 32 with total number 30, diversity=2, DIA and auxiliary criterion support 
\addplot[
color=violet,
mark=triangle,
]
coordinates {
(2.2,0.01171)
(2.4,0.00658)
(2.6,0.00354)
(2.8,0.00178)
(3.0,0.0009)
(3.2,0.00039)
};	
\addlegendentry{NMS-OSD($G_r$=30)}

%plot for NMS-1,T=13, OSD of order-pattern size 32 with total number 100, diversity=2, DIA and auxiliary criterion support 
\addplot[
color=red,
mark=triangle*,
]
coordinates {
(2.2, 0.00809) 
(2.4, 0.00439) 
(2.6, 0.0021) 
(2.8, 0.00112) 
(3.0, 0.00049) 
(3.2, 0.00021)
};	
\addlegendentry{NMS-OSD($G_r$=100)}

%plot for NMS-1,T=13, OSD of order-pattern size 32 with total number 200, diversity=2, DIA and auxiliary criterion support 
\addplot[
color=purple,
mark=diamond,
]
coordinates {
(2.2, 0.00679) 
(2.4, 0.00355) 
(2.6, 0.00154)
(2.8, 0.00082) 
(3.0, 0.00034) 
(3.2, 0.00013)
};
\addlegendentry{NMS-OSD($G_r$=200)}

%plot for D_{10}$-OSD-2(25)
\addplot[
color=orange,
mark=diamond*,
very thin
]
coordinates {
  (2.50, 2.2e-03)
  (3.00, 3.0e-04)
  (3.25, 9.0e-05)
  (3.5,2.5e-5)
};	
\addlegendentry{$D_{10}$-OSD-2(25) \cite{Rosseel2022}}

%plot for ML (FER)
\addplot[
color=cyan,
mark=asterisk,
very thin
]
coordinates {
(1.00, 1.064e-01)
(1.50, 3.397e-02)
(2.00, 8.773e-03)
(2.50, 1.168e-03)
(3.00, 1.321e-04)
(3.50, 1.022e-05)
};	
\addlegendentry{ML \cite{helmling19}} 

\end{semilogyaxis}
\end{tikzpicture}
\caption{FER comparison for various decoding schemes of CCSDS (128,64) code}
\label{fer_128_diversify}
\end{figure}

%% file: plots/fer_384_NMS_diversify.tex
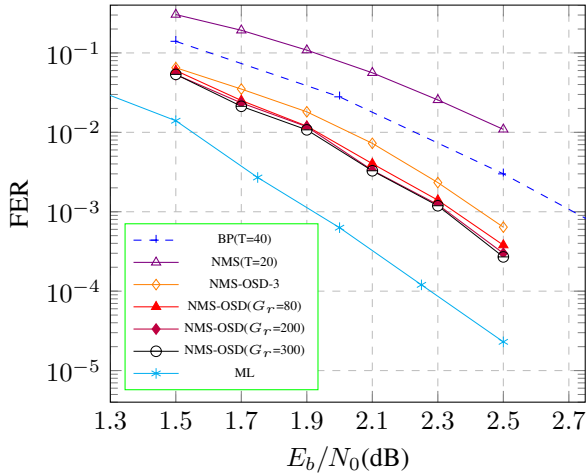
\begin{figure}[htbp]
\centering
\begin{tikzpicture}
\begin{semilogyaxis}[
    scale = 0.75,
    xlabel={$E_b/N_0$(dB)},
    ylabel={FER},
    xmin=1.3, xmax=2.75,
    ymin=4e-6, ymax=4e-1,
    xtick={1.3,1.5,1.7,1.9,2.1,...,2.8},
    legend pos = south west,
    ymajorgrids=true,
    xmajorgrids=true,
    grid style=dashed,
    legend style={font=\tiny},
    ]
%1 plot for BP (FER)
\addplot[
color=blue,
mark= +,
dashed,
]
coordinates {
(1.5,1.4e-1)
(2.0,2.8e-2)
(2.5,3e-3)
(3.0,2.3e-4)
};	
\addlegendentry{BP(T=40)}
%plot for NMS-2(T=20)
(1.5,0.3043),(1.7,0.1928),(1.9,0.1084),(2.1,0.0562),(2.3,0.0256),(2.5,0.0109)

\addplot[
color=violet,
mark=triangle,
]
coordinates {
(1.5,0.3043)
(1.7,0.1928)
(1.9,0.1084)
(2.1,0.0562)
(2.3,0.0256)
(2.5,0.0109)
};	
\addlegendentry{NMS(T=20)}
%plot for NMS-2 plus conventional OSD-3
\addplot[
color=orange,
mark=diamond,
very thin
]
coordinates {
(1.5,0.06511)
(1.7,0.03508)
(1.9,0.01813)
(2.1,0.00727)
(2.3,0.00233)
(2.5,0.00064)
};	
\addlegendentry{NMS-OSD-3}

%plot for NMS-2,T=20, OSD of order-pattern size 32 with total number 80, diversity=3, DIA and auxiliary criterion support 
\addplot[
color=red,
mark=triangle*,
]
coordinates {
(1.5,0.06002)
(1.7,0.02518)
(1.9,0.01197)
(2.1,0.00403)
(2.3,0.0014)
(2.5,0.00038)
};	
\addlegendentry{NMS-OSD($G_r$=80)}

%plot for NMS-2,T=20, OSD of order-pattern size 32 with total number 200, diversity=3, DIA and auxiliary criterion support 
\addplot[
color=purple,
mark=diamond*,
]
coordinates {
(1.5,0.05339)
(1.7,0.02346)
(1.9,0.01178)
(2.1,0.00333)
(2.3,0.00123)
(2.5,0.0003)
};
\addlegendentry{NMS-OSD($G_r$=200)}

%3 plot for NMS-OSD($G_r$=300) FER	
\addplot[
color=black,
mark=halfcircle,
very thin
]
coordinates {
(1.5,0.05367)
(1.7,0.02128)
(1.9,0.01075)
(2.1,0.00328)
(2.3,0.00119)
(2.5,0.00027)
};
\addlegendentry{ NMS-OSD($G_r$=300)}
%plot for ML (FER)
\addplot[
color=cyan,
mark=asterisk,
very thin
]
coordinates {
(1.25, 3.5e-02)
(1.5,1.4e-2)
(1.75, 2.7e-03)
(2.00, 6.3e-04)
(2.25, 1.2e-04)
(2.5, 2.3e-05)
};	
\addlegendentry{ML}

\end{semilogyaxis}
\end{tikzpicture}
\caption{FER comparison for various decoding schemes of Wimax-like (384,192) code}
\label{fer_384_diversify}
\end{figure}

%% file: plots/fer_1056_NMS_dynamics.tex
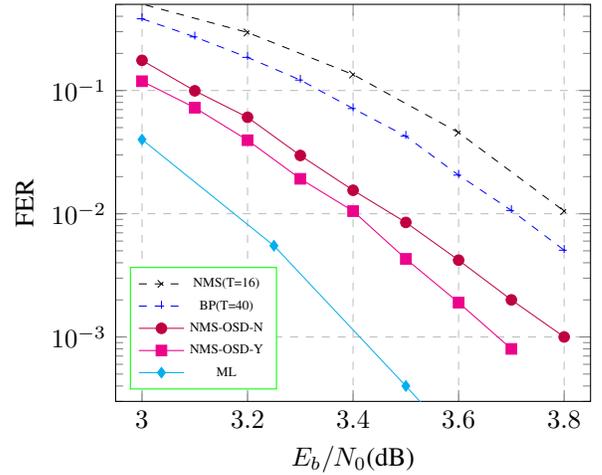
\begin{figure}[htbp]
	\centering
	\begin{tikzpicture}
		\begin{semilogyaxis}[
			%title=FER comparison of various decoding schemes for  code (1056,880)},
			scale = 0.75,
			xlabel={$E_b/N_0$(dB)},
			ylabel={FER},
			xmin=2.95, xmax=3.85,
			ymin=3e-4, ymax=5e-1,
			xtick={3.0,3.2,3.4,...,4.},
			legend pos = south west,
			ymajorgrids=true,
			xmajorgrids=true,
			grid style=dashed,
			legend style={font=\tiny},
			]

%0 plot for NMS-2(16)
\addplot[
color=black,
mark=x,
dashed,
very thin
]
coordinates {
(2.8,0.7167)
(3.0,0.5025)
(3.2,0.2957)
(3.4,0.1347)
(3.6,0.0455)
(3.8,0.0105)
};
\addlegendentry{NMS(T=16)}
%3plot for BP(40) FER				
\addplot[
color=blue,
mark= +,
dashed,
very thin
]
coordinates {
(3.0,0.381875)
(3.1,0.2720192307692308)
(3.2,0.18566176470588236)
(3.3,0.12151041666666666)
(3.4,0.07140116279069768)
(3.5,0.042535714285714284)
(3.6,0.02056451612903226)
(3.7,0.010627104377104377)
(3.8,0.005058061420345489)
};
\addlegendentry{BP(T=40)}

%5 plot for NNMS(20)-DIA-OSD(6)
\addplot[
color=purple,
mark=*,
very thin
]
coordinates {
(3.0, 0.1757) 
(3.1, 0.0994) 
(3.2, 0.0606) 
(3.3, 0.0297) 
(3.4, 0.0155) 
(3.5, 0.0085) 
(3.6,0.0042)
(3.7,0.002)
(3.8,0.001)
};
\addlegendentry{NMS-OSD-N}
%5 plot for NMS-2-OSD(D-3)-Y
\addplot[
color=magenta,
mark=square*,
]
coordinates {
(3.0, 0.1189)
(3.1, 0.0725) 
(3.2, 0.0395) 
(3.3, 0.0192) 
(3.4, 0.0105) 
(3.5,0.0043) 
(3.6,0.0019)
(3.7,0.0008)
};
\addlegendentry{NMS-OSD-Y}

%0.5plot for ML (FER)
\addplot[
color=cyan,
mark=diamond*,
very thin
]
coordinates {
(3, 4.0e-2)
(3.25, 5.5e-3)
(3.50, 4.0e-04)
(3.75,2.7e-5)
};	
\addlegendentry{ML}
\end{semilogyaxis}
\end{tikzpicture}
\caption{Performance comparison for various decoding schemes of 802.16 (1056,880) code}
\label{fer_1056}
\end{figure}

%% file: sections/conclusions.tex
\section{Conclusions and Future Research}
\label{conclusions} 

This paper initiates an exploration into the design motivation of the NMS family, primarily from the perspective of graphical neural networks. Our investigation reveals that the simplest one, characterized by a sole parameter, is preferred due to its optimal balance between decoding performance and complexity. In the adaptive OSD component of the NMS joining OSD framework, we recommend either uniformly or dynamically partitioning the TEPs to populate order patterns, depending on the code block lengths. Simultaneously, iteration diversity is implemented by splitting the iterative decoding trajectories of the decoding failures into multiple groups, with the benefits of not only facilitating the reduction of complexity in the DIA model but also contributing to the improvement of OSD performance. Empirical verification demonstrates that the decoding performance of our adopted architecture can readily approach that of the state-of-the-art decoder with significantly fewer computational resources.

It is noteworthy that the findings regarding NMS selection in this paper align with those of \cite{lian19}, where the authors compared all NBP and NMS variants for BCH codes from the perspective of neural network sharing or tying techniques, reaching a conclusion similar to ours in the context of graphical neural networks.

Intuitively, exploring multiple loss definitions to achieve diversity gain for the OSD is a promising direction for future research. Additionally, for longer codes, investigating more intricate schemes like DIA to fully leverage the iterative decoding trajectories remains an open area. Improving the bit error rate (BER) metric for the OSD decoding is also an interesting avenue for future exploration.

\section*{Availability of Data}
Parity check matrices of the three codes under discussion, as well as BP and ML decoding results, can be accessed on the website \cite{helmling19}, thanks to the generosity of the original authors and the dedicated efforts taken for its maintenance.